# Improved Explanatory Efficacy on Human Affect and Workload Through Interactive Process in Artificial Intelligence

**BYUNG HYUNG KIM, SEUNGHUN KOH, SEJOON HUH, SUNGHO JO, AND SUNGHEE CHOI**
School of Computing, KAIST, Daejeon 34141, South Korea

Corresponding author: Sunghee Choi (sunghee@kaist.ac.kr)

This work was supported by Institute of Information & Communications Technology Planning & Evaluation (IITP) grant funded by the Korea government (MSIT) (No. 2017-0-00432).

**ABSTRACT** Despite recent advances in the field of explainable artificial intelligence systems, a concrete quantitative measure for evaluating the usability of such systems is nonexistent. Ensuring the success of an explanatory interface in interacting with users requires a cyclic, symbiotic relationship between human and artificial intelligence. We, therefore, propose explanatory efficacy, a novel metric for evaluating the strength of the cyclic relationship the interface exhibits. Furthermore, in a user study, we evaluated the perceived affect and workload and recorded the EEG signals of our participants as they interacted with our custom-built, iterative explanatory interface to build personalized recommendation systems. We found that systems for perceptually driven iterative tasks with greater explanatory efficacy are characterized by statistically significant hemispheric differences in neural signals with 62.4% accuracy, indicating the feasibility of neural correlates as a measure of explanatory efficacy. These findings are beneficial for researchers who aim to study the circular ecosystem of the human-artificial intelligence partnership.

## I. INTRODUCTION

Recent advances in artificial intelligence (AI) and machine learning algorithms have resulted in models that not only achieve high predictive performance but also provide explanatory features to support their decisions, increasing model interpretability and transparency in real-world environments [1].

However, merely providing explanations is insufficient. Ultimately, AI should address the problems hindering human-agent interaction. Much of the current work for human-interpretable machine learning systems suffers from a lack of usability and efficacy [2]. Particularly, there have been limitations in developing interface technology that reflects user feedback onto machine systems due to the lack of quantitative evaluation of how the user perceives the explanations to be. Failing to integrate user knowledge with machine systems can decrease interaction quality to the point of causing interaction breakdowns. Consequently, the systems will lose their ability to justify their recommendations, decisions, or actions, resulting in a loss of trust from their users. Therefore, ensuring the success of an explanatory interface in interacting with users requires a cyclic, symbiotic relationship (FIGURE 1). The system should provide explanations about its decision-making process, giving users some insight into how the system will behave. By hypothesis, explanations that are succinct and easily interpretable should enable users to develop appropriate trust in the AI and perform well when using the AI. Users can subsequently increase both their understanding and the model's performance by correcting what they perceive to be the system's flawed reasoning [3]. The system, in turn, reflects the feedback by changing how it makes decisions, and the cycle continues. Accordingly, developing such a feedback-based interface for explainable AI (XAI) systems requires an evaluation on the strength of the cyclic relationship the interface exhibits, which is defined as explanatory efficacy in our work. Although many different types of metrics for evaluating interface technologies have been investigated and highlighted for several theoretical and empirical works [4], [5], they lack insight into the user's affective-cognitive processes. Psychological research suggests that evaluating performance without considering affective-cognitive processes may lead to incorrect conclusions about the efficacy of an interface [6], [7]. Moreover, as the explanatory interfaces support analytical reasoning tasks, its evaluation should consider the emotional influence and cognitive demand of human-AI interaction [8].

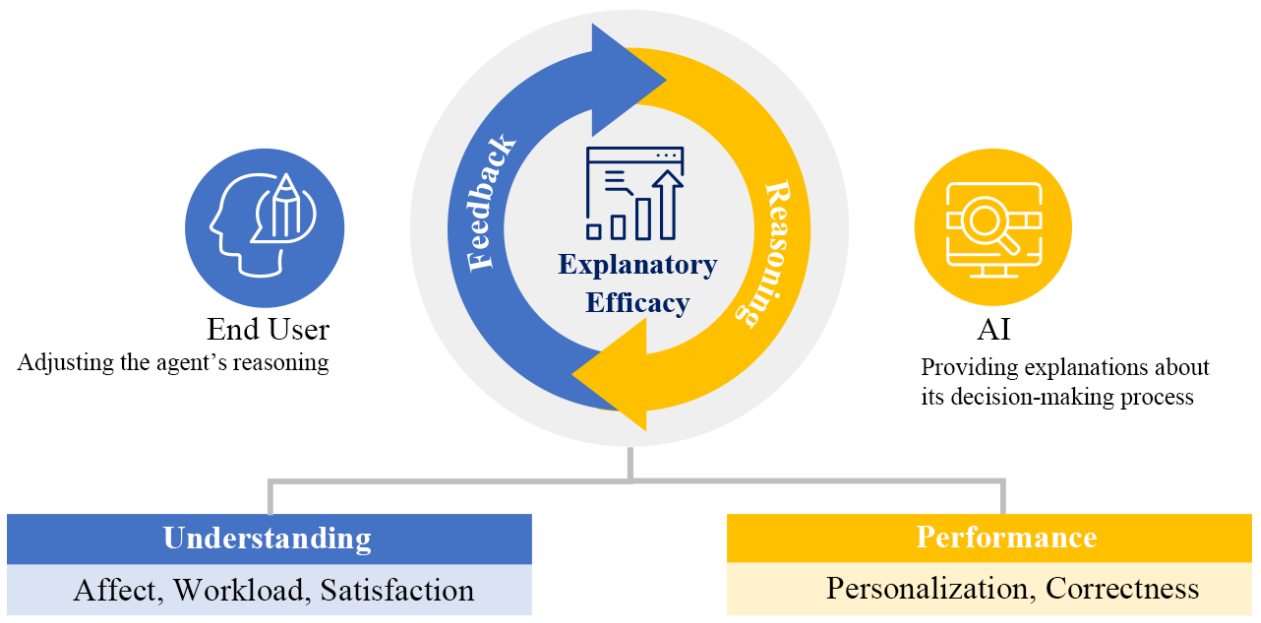

**FIGURE 1. Ensuring the success of an explanatory interface in interacting with users requires a cyclic, symbiotic relationship. Developing such a feedback-based interface for AI systems requires an evaluation on the strength of the cyclic relationship the interface exhibits, which we define as *explanatory efficacy*.**

One possible method of studying affective-cognitive processes is by using electroencephalography (EEG). During the recent decade, EEG has received increased attention as a lightweight sensing technology for recording brain activities. Its exceptionally high temporal resolution is valuable in studying neurophysiological phenomena in affective-cognitive processes. Furthermore, the non-invasiveness and mobility of EEG have extended its usage to the field of brain-computer interfaces (BCIs), external devices that communicate with the user's brain [9]. Nevertheless, previous work has only focused on investigating visual tasks or user workload [7], [10], and has not been used to measure the explanatory efficacy of interactive machine systems for iterative tasks.

Hence, in this work, we aim to explore explanatory efficacy through interactive process in an XAI system and analyze the effect on human affect and workload through EEG signals. Furthermore, we demonstrate the potentiality of neural correlates in EEG signals as a new measure of explanatory efficacy. We believe that the implemented XAI system in this study enables users to understand, appropriately trust, and effectively manage the system by helping them comprehend the rationale behind the system's decisions. Prior work has focused on how an intelligent agent can explain itself to end-users [11], and how end-users may act upon such predefined explanations by human experts to debug their intelligent agents [4], [12]. This work, in contrast, considers how a machine-generated explanation and a user's feedback can impact the user's affective-cognitive processes associated with explanatory efficacy.

With explanatory efficacy in mind, we further explore the potentiality of neural correlates in EEG signals as a measure of explanatory efficacy unlike previous works that have studied the physiological effect of user's behaviors on interface technologies. Toward this end, we investigated three research questions:

1) Feasibility (Q1): Can the explanatory efficacy of an interactive XAI system's recommendation be improved by feedback (a user correcting what they perceive to be the system's flawed reasoning)?
2) Experience (Q2): Do end-users experience differences in terms of human affect and workload when interacting with a system of high explanatory efficacy than with a system of low explanatory efficacy?
3) Potentiality (Q3): Can the physiological characteristics measured by EEG be used to evaluate explanatory efficacy?

To answer these questions, we conducted an empirical study that monitored the user's affective-cognitive processes through EEG signals. Simultaneously, we investigated the effects of explaining the reasoning of a new personalized movie recommendation system with explanatory features to a user and allowing for the user to personalize their recommendations through interactions with the system.

Our findings indicate that the explanatory efficacy of an interface can be evaluated with EEG signals associated with human affect and workload. Particularly, we observed that the physiological characteristics of EEG signals correlate with human affect and workload in perceptually driven iterative tasks and that an increased explanatory efficacy can lead to improvements in the model's ability to predict personalized results.

### A. PROBLEM STATEMENT: EXPLANATORY EFFICACY

Given an interactive trial $t$, suppose an XAI system provides a set of prediction $P_t(m) = (p_1, p_2, \ldots, p_M)$ along with a sequence of explanation $E_t(m, n) = (e_1^1, e_2^1, \ldots, e_1^2, e_2^2, \ldots, e_N^M)$ about its decision-making process. A successful cyclic relationship should result in end-users deepening their understanding of the machine's explanation and the machine system increasing predictive performance. Hence, measuring explanatory efficacy $\xi_t$ is to simultaneously evaluate users' model understanding and the system's predictive performance.

$$\xi_t = a_t f(x_t), \tag{1}$$

where $f(\cdot)$ is a family of linear functions, $a_t = \sum_{m=1}^{M} S(P_t(m))$, and $x_t = \sum_{m=1}^{M} \sum_{n=1}^{N} U_t(E_t(m, n))$. $S_t(\cdot)$ is the level of the system's perdictive performance. $U_t(\cdot)$ is the depth of the user's understanding. $M$ and $N$ are the number of predictive results and the number of explanation, respectively. We omit $t$ for simplicity.

To evaluate how effectively the explanatory process in the cyclic relationship delivered explanations to end-users, their thoughts on the interface's explanatory efficacy for each iteration should be quantified. Each of the questions asked for each iteration $t$ measured the depth of understanding for a specific explanatory feature $e_n^m$ of the system's prediction $p_m$, and their combination $x = \sum_{m=1}^{M} \sum_{n=1}^{N} U(E(m, n))$ serves as a reasonable proxy for participants' understanding of the entire system. We calculated the depth of understanding $U(\cdot)$ as follows:

$$U(E(m, n)) = C_r(E(m, n)) \times C_f(E(m, n)), \tag{2}$$

where $C_r(\cdot)$ is either 1 for a correct explanation, or -1 for an incorrect explanation. $C_f(\cdot)$ is a confidence value between 1 and 9. These values were summed for each explanatory feature $E(m, n)$ to create a user's understanding score.

To evaluate how the machine's decision-making process improved its performance within the cyclic relationship, the performance $S(\cdot)$ can be measured as follows:

$$S(P(m)) = \begin{cases} 1 & \text{if } P(m) \text{ is correct} \\ 0 & \text{else} \end{cases} \tag{3}$$

## II. RELATED WORKS

### A. EXPLAINABLE ARTIFICIAL INTELLIGENCE (XAI) SYSTEMS

Many recent studies in XAI have explored the potential of AI by building more transparent, explainable, or interactive systems so that users can be better equipped to understand and therefore trust the intelligent systems [1], [13]. Such studies include supporting autonomous agent behavior [14], explaining predictions of multiple classifiers or machine's decision-making [15], and debugging machine learning systems [4]. Hayes and Shah [14] presented an explainable mechanism to better calibrate the expectations of the machine's behaviors. To explain control policies, they produced behavioral explanations in such a way that humans could understand them. Ribeiro *et al.* [15] proposed a system that explains the predictions of any classifier for text and image classification by learning an interpretable model locally around the prediction. While the aforementioned studies do increase the trustworthiness and transparency of intelligent agents, from the perspective of end-users, these works suffer from a lack of usability, practical interpretability, and efficacy. Although current research in XAI has focused on developing 'more AI' systems, we believe XAI will ultimately be a problem of human-agent interaction, which can be defined as the transdisciplinary field of AI, social science, and human-computer interaction (HCI).

### B. EXPLAINABILITY IN ARTIFICIAL INTELLIGENCE SYSTEMS

As an increasing number of human-facing applications utilize XAI, there has been an increase in designing interfaces for humans to interpret and interact with those systems and their predictions [2]. In response, researchers have proposed a variety of novel interfaces for interacting with machine learning models, but these tend to require the users to have significant expertise in specific tasks [12], [16], [17]. However, developing an explanatory interface in which the system's decisions are elucidated by machine-generated explanations is especially challenging, as the interface needs to evaluate the auto-generated information while taking into consideration the user's understanding of dynamical and complex systems [4].

Mental models are internal representations of how people empirically understand phenomena [18]. They can be constructed from perception, imagination or the comprehension of decision-making systems. Empirical studies on mental models have shown that users may build their own mental models and change them when an intelligent system makes its reasoning transparent [11], [19]. Furthermore, explicit instructions regarding new features of an intelligent machine system, such as why…and why not…descriptions of the agent's reasoning, can also improve mental models as they increase the system's transparency [20]. However, increasing transparency of reasoning in intelligent systems does not ensure that users will build mental models of higher quality. The link between end-users' mental models and their satisfaction with the intelligent system's behavior has not yet been fully investigated. For example, the satisfaction levels of experienced users may actually decrease as a result of more transparency [19]. Furthermore, evaluation of mental models requires knowledge on how models of human affect and cognition are elicited and represented. Hence, in this article, we aim to study neural correlates of explanatory efficacy with human affect and cognitive workload during interactions with the system.

### C. AFFECTIVE-COGNITIVE PROCESS IN EXPLANATION EVALUATION

Human-XAI interaction is a complex affective-cognitive system since they are associated with individual beliefs. To our knowledge, there have been only two significant attempts in estimating the affective-cognitive process in interface technologies: questionnaires and physiological sensing.

Recording behavioral metrics and administering questionnaires represent discrete and sporadic events that reflect aggregated opinions about a whole experience [4], [11]. On the other hand, physiological sensing has the advantage of having higher temporal fidelity in that it can access data at any time [7], [21]. Empirical studies on physiological sensing have shown how human mentality can be represented in the affective-cognitive process with an underlying theoretical foundation in neuroscience [22]. Particularly, valence and motivational engagement, two of the various dimensions of the human affect, have been investigated and used to measure the level of perception in human-AI interaction [22], [23]. For example, the valence hypothesis states positive emotions are processed by the left hemisphere and negative emotions by the right hemisphere. On the other hand, the approach-withdrawal hypothesis states approach emotions (motivational engagement) are processed by the left hemisphere and withdrawal emotions by the right hemisphere. Their relative simplicity on the asymmetric measure has been advantageous

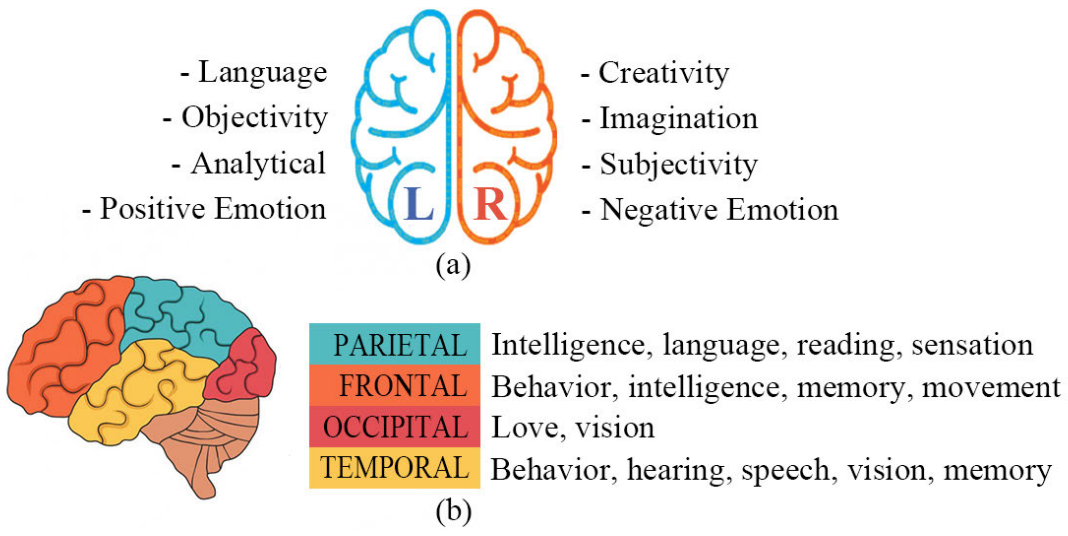

**FIGURE 2.** **Brain structure. (a) The cerebral cortex includes the frontal, temporal, occipital and parietal lobes. (b) Left analytical and right creative hemispheres.**

for computational models when it comes to recognizing affective and cognitive states in the brain.

### D. EEG FOR AFFECTIVE-COGNITIVE EVALUATION

EEG is a measure of electrical waveforms from the scalp that is used to analyze cortical electrical activity (FIGURE 2). Human EEG waves on the frontal, temporal, parietal, and occipital lobes have four main frequencies (theta, alpha, beta, gamma), all of which have been used to study neural activity in various research areas [24]. Particularly, research on the neuroscientific characteristics of the affective-cognitive processes have focused on investigating the alpha and beta bands on the frontal and temporal lobes since they are responsible for conscious thought, perception, emotions, and personality [25]. Kim and Jo [22] explored how EEG features extracted from different lobes are correlated with emotional changes. They discovered that increased theta and alpha power over the frontal regions led to an increase in valence and arousal. The high temporal resolution of EEG has led to its widespread usage in many clinical applications. For instance, by detecting temporal abnormalities in EEG signals, epilepsy and other sleep disorders can be identified [26], [27]. Therefore, in this study, we utilize EEG signals for enhancing the usability of interactive machine learning.

## III. XAI-BASED RECOMMENDATION SYSTEM

In order to investigate our work's research questions (Q1, Q2, and Q3), we implemented an XAI-based movie recommendation system in which users get personalized movie lists that fit their particular tastes. The main goal of the interface is to realize transparency; users should be able to understand how the system makes its predictions and thereby know when to trust the system's decisions. Users can also refer to explanations of the system's decisions and provide their feedback by adjusting the agent's reasoning so that future recommendations more closely match the user's desired types of movies.

### A. PREDICTIVE MODEL

The movie recommendation system uses a retrieved list of the movie's attributes such as names of actors and directors, genres, and tag data as explanatory features. Each of these features are given a weight from 0 to 100 that signifies the extent to which that feature positively affected the user's rating of the selected movie. The interface utilizes visual aids in the form of sliders and bar diagrams to emphasize and convey the weights of each feature to the user, helping the user understand the reasoning underlying the overall estimate of the user's movie preferences.

FIGURE 3 shows an overview of our proposed interface and the screenshot of the main result. User interactions with our system proceed as follows. Before each interaction, the interface initially recommends 20 movies that the system predicts the user will like the most. Next to each of these movies, the three explanatory features that influenced the prediction the most and their corresponding weights are also displayed (FIGURE 3a). For further transparency, the interface allows for users to click on a movie for more details, which include a thorough list of all relevant features and their corresponding weights, so that the user can better understand the model's recommendation.

Along with the additional information that is provided when a movie is clicked, the interface also displays the six most influential explanatory features and their corresponding weights (FIGURE 3b), which are presented as sliders, separately. This is a key feature of our interface; users can then provide their feedback to the system by moving the sliders to alter the respective weights. With the new weights, the system makes new predictions and recommends a new list of movies for the user. Users can then directly observe how their feedback affected the new list of recommendations (FIGURE 3c). Users can also express their satisfaction for the recommendation by clicking either the 'Like' or 'Dislike' button.

This interactive feature is beneficial in two ways: 1) it increases transparency as users can see how model predictions about personal preference contribute towards the overall list of movie recommendations, and 2) users can directly contribute to a more personalized prediction by changing the weights of some features that they disagree with. For example, one user may enjoy the movie Deadpool because of his interest in superhero movies. However, the movie recommendation system may have the "Horror" genre feature to be one of the top six influential features for why they recommended the movie. The user may then lower the weight of the "Horror" genre feature so that the system knows the feature did not positively affect the user's viewing experience as much as the system thought it did.

In summary, for every iteration, the user receives a list of movie recommendations and provides feedback by changing the weights of a specific recommendation. Based on the feedback, a more personalized recommendation is made and presented to the user.

### B. PREDICTION WITH INTERACTION

Our system's automated predictions are based on deep learning-based clustering algorithms. The models are trained with the Movielens dataset[1] and the IMDB website's dataset.[2] In total, from the two datasets, we retrieved the unique ID, title, genres, tags, and ratings of 55,846 movies.

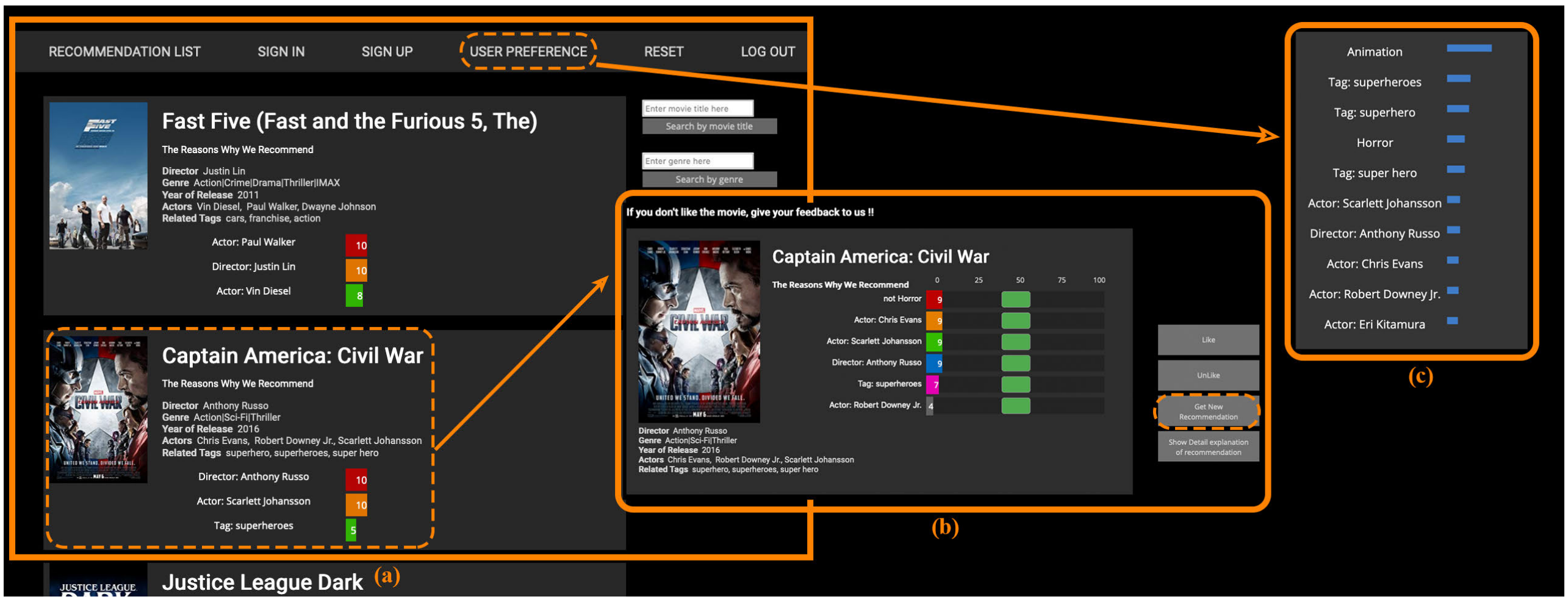

**FIGURE 3.** **The overview of our proposed interface and snapshot of the main result. Clicking on the movie goes to a new page where the top 6 features for why Captain America was recommended to the user is provided. The user can then adjust these weights if the user thinks they are incorrect, and this feedback is subsequently used to update the weights of the explanatory features and provide new recommendations. (a) The main display of 20 recommended movies, each with the three most influential explanatory features and their corresponding weights. (b) A clicked movie ('Captain America: Civil War') with the six most influential explanatory features. Users can provide *feedback* to the system by moving the weight sliders. They can also express whether they are satisfied with recommendations by clicking either the 'Like' or 'Dislike' button. (c) Personalized prediction of the user's movie preferences.**

For inference and learning, the system builds the initial movie recommendation list by clustering similar movies based on genre and tag features using an autoencoder [28] with the ADAM learning algorithm [29]. The recommendation list is then further personalized in every subsequent iteration by reflecting users' feedback on their movie preferences. Whenever users want new recommendations by pressing the 'Get new recommendation' button, the predictive model retrieves a pool of 40 movies that have similar features to movies previously seen by the users. Given the pool, a decision-tree regression model is used with a maximum depth of 40 for training. Note that at least three data points are required for a leaf to split. Then, the regression model produces the expected ratings that it predicts the user will give to each movie. The system finally chooses and displays the 20 movies with the highest expected ratings.

The underlying model architecture, which uses these predictions to estimate the user's preference in movies, is designed to improve transparency. To provide explanations for the system's predictions, we analyze the user-specific data along with the 20 chosen movies using Local Interpretable Model-agnostic Explanations (LIME), which is a widely used algorithm to explain model predictions. It can provide insight into each decision by locally approximating the classifiers [15]. In our work, LIME produces the weights of each explanatory feature with which the decision-tree regression model predicts the ratings of the movies.

[1]https://www.movielens.org

[2]https://www.imdb.com

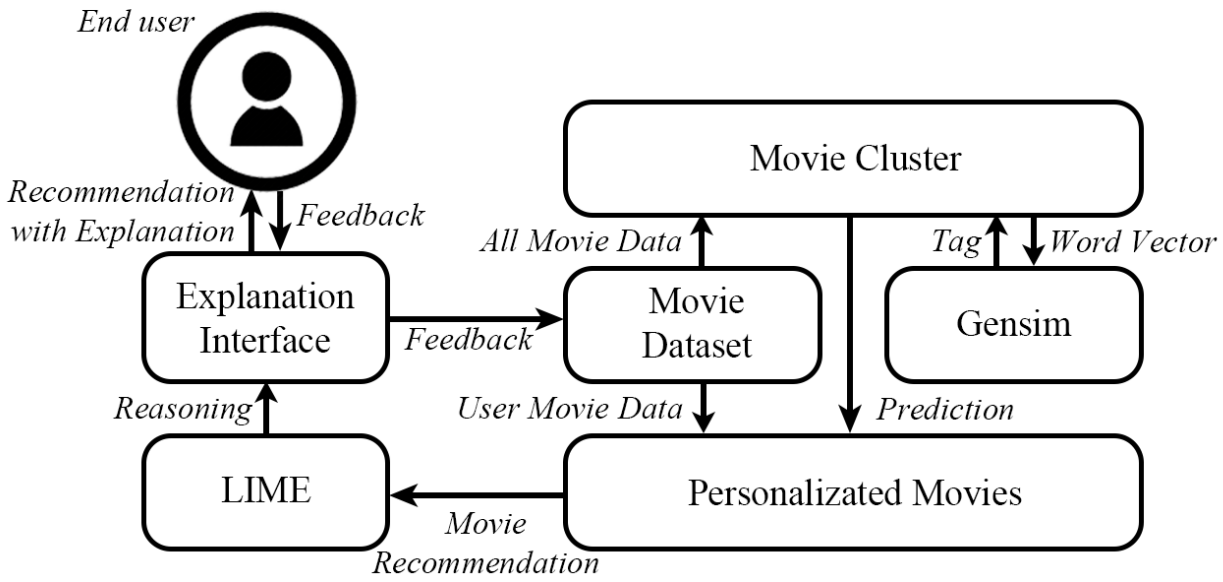

**FIGURE 4.** **The overview of the predictive model for movie recommendations.**

Changing the weights causes a change in the expected rating of a movie with the changed feature. This change, denoted as $\hat{y}$, is computed with the equation below:

$$\hat{y} = c \times r \times \frac{\omega_a - \omega_b}{100}, \tag{4}$$

where $\omega_b$ and $\omega_a$ are the weight of the chosen feature before and after the user changes it, respectively. $r$ is the currently expected rating of the movie and $c$ is a constant.

The value of $c$ depends on the type of the feature that the user changed. If the feature is genre-related, then $c$ is 0.15, but otherwise, it is 0.3. Because genre occupies 20 out of 28 dimensions of the input data, a change in the weight of a genre-related feature changes the model much more significantly than does the same change in the weight of any of the other features. To regulate such a behavior, the value of $c$, when the feature is genre-related, was reduced to 0.15 through trial and error. Then, the value of $\hat{y}$ is added to the expected rating of every movie with the feature whose weight was changed. When the regression model is trained with the user's updated personal data, the regression model recognizes that the user changed the weight of a specific feature and recommends accordingly with the changed tendency.

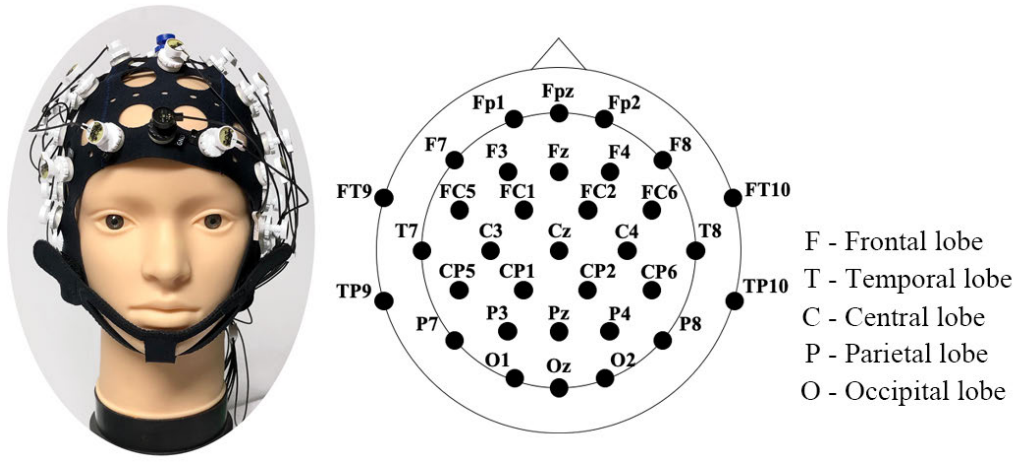

**FIGURE 5.** **The EEG device with 32 active electrodes (Left) and the electronode placement (frontal, temporal, central, parietal, and occipital lobes) according to the international 10-20 system (Right).**

## IV. USER STUDY

We recruited 24 participants (20 male and 4 female) between the ages of 21 and 38 for the study. An incentive of \$20 was given to each participant. The study was conducted in a lab setting, with the proposed interface presented on a single monitor. Participants first had a tutorial session about the functionality of our interface. They were asked to search for any 20 movies that wanted to. For the first movie they searched, the weights of the features were at their default values. From there, users were asked to click either the 'Like' or 'Dislike' button to express their satisfaction with the system's explanations. Clicking the 'Dislike' button reduces the weights of all the features by half. Clicking the 'Like' button does not change the weights, as the user is satisfied with how the system is predicting and explaining their recommendation. These 20 movies were used for initializing the recommendation system. After the tutorial, participants were allowed to do as many iterations as they wished (averaged $18.2 \pm 3.4$ iterations) by pressing the 'Get new recommendation' button to move onto the next iteration. More precisely, users continued to iterate through the interface until they felt it unnecessary to continue to update the weights due to either a great satisfaction or dissatisfaction with the system. After completing the study, participants completed a short survey.

We recorded EEG signals using a Brain Products system[3] in a laboratory environment. The EEG signals were recorded at a sampling rate of 500 Hz on 32 active AgCl electrodes placed according to the international 10-20 system (FIGURE 5). The EEG data were common average referenced, downsampled to 250 Hz, and high-pass filtered with a 2 Hz cutoff frequency. After removing eye artifacts with a blind source separation technique[4] (independent component analysis [30] with manually selected three components), we extracted the EEG signals starting from 30 seconds before clicking the 'Get new recommendation' button until 30 seconds after clicking it [22]. The EEG signals from the 5 seconds in the middle of the interval between each click of the 'Get Recommendation Button' was extracted as a baseline to correct for unrelated variations in power over time [31].

[3]https://www.brainproducts.com

[4]https://www.martinos.org/mne

The frequency power of interactive trials and the baseline between 3 and 47 Hz was extracted with Welch's method with 50% overlap and 250 DFT length. The baseline power was then subtracted from the feedback period power, yielding the change of power relative to the non-feedback period. These changes of power were averaged over the frequency bands of theta (4 - 7Hz), alpha (8 - 13Hz), beta (14 - 29Hz), and gamma (30 - 47Hz). This study was approved by the KAIST Institutional Review Board (IRB) in Human Subjects Research. All research was performed in accordance with the relevant guidelines and regulations. Informed consent was obtained from all participants.

### A. PROCEDURE

Participants were randomly assigned to one of two groups. The participants in the *non-feedback* explanatory interface group (9 male and 3 female) were not allowed to give any feedback by changing the weights; they could only click either the 'Like' or 'Dislike' button. On the other hand, those who were in the *feedback* explanatory interface group (11 male and 1 female) could provide feedback not only through the 'Like' and 'Dislike' buttons but also by changing the features' explanatory weights so that the system would better reflect the user's preferences. Clicking the 'Dislike' button, unbeknownst to the participants, reduces the weights of all the features by half unless they were in the *feedback* group and specified their own changes to the explanatory weights. We logged all key-strokes and response times and recorded EEG signals for all participants as they used the proposed interface. At the end of each iteration, all participants completed unweighted affective-cognitive questionnaires. Each interactive trial consists of the following steps:

1) Model prediction : 20 movie recommendation (FIGURE 3(a)).
2) Model explanation : 10 explanations (FIGURE 3(c)).
3) Explanatory efficacy and affective-cognitive assessment.
4) User feedback on the model prediction and 6 explanation (FIGURE 3(b)).
5) New recommendation by pressing the 'Get new recommendation' button.
    - EEG signals starting from 30 seconds before clicking the button until 30 seconds after clicking it.

### B. MEASURE OF EXPLANATORY EFFICACY

To evaluate $U(\cdot)$ in Eq.(2), we asked participants about their thoughts on the interface's explanatory efficacy for each iteration $t$. Each of the questions asked for each iteration measured the depth of understanding for a specific explanatory feature (FIGURE 3c) of the system's predictive recommendation. Their combination $x$ in Eq.(1) serves as a reasonable proxy for participants' understanding of the entire system, ranging from $-90$ (indicating a participant who was completely confident that every explanation was wrong) to $+90$ (indicating a participant who was completely confident that every explanation was correct) from the 10 explanatory features (FIGURE 3c) in our study. To evaluate $S(\cdot)$ in Eq.(3), participants also expressed whether they were satisfied with recommendations by clicking either the 'Like' or 'Dislike' button throughout the study. Since satisfaction about the recommendation reflects the reliability of the system's performance, we used the number of 'Likes' as a metric for model performance on personalization. Lastly, we used a logistic function for $f(\cdot)$ in Eq.(1) as follows:

$$f(x) = \frac{1}{1 + e^{-kx}}, \tag{5}$$

where $x = \sum_{m=1}^{M} \sum_{n=1}^{N} U(E(m, n))$ and $k = 90$ for stabilizing the function.

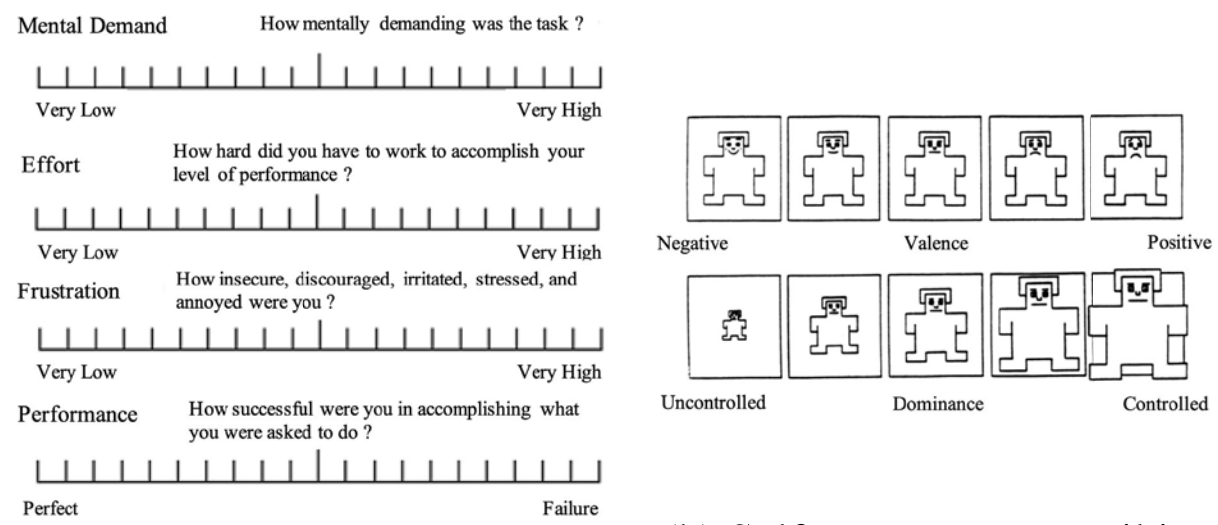

(b) Self-assessment manikins

(a) NASA-TLX

**FIGURE 6.** **Self-assessment of the levels of cognitive and affective states.**

### C. PARTICIPANT SELF-ASSESSMENT

At the end of each iteration, participants performed a self-assessment of their level of cognitive (mental demand, performance, effort, and frustration) and affective (valence and dominance) states. Unweighted NASA-TLX questionnaire [32] and Self-assessment manikins [33] were used to visualize the scales (FIGURE 6). Particularly, the manikins were displayed in the middle of the screen with the numbers 1-9 printed below. Participants moved the mouse strictly horizontally just below the numbers and clicked to indicate their self-assessment level. Participants were informed they could click anywhere directly below or in-between the numbers, making the self-assessment a continuous scale. The valence scale ranges from unhappy or sad to happy or joyful. The dominance scale ranges from submissive (or "without control") to dominant (or "incontrol, empowered").

## V. FEASIBILITY OF EXPLANATORY EFFICACY

Our feedback-based explanation interface provides the reasoning behind the system's decisions and receives feedback from the user, which allows for the system to reflect the feedback in future decisions. Hence, this study analyzes whether the implementation and usage of such a feedback-based interactive system improves explanatory efficacy, increasing the users' understanding and satisfaction of the model's recommendation.

We tested for differences in explanatory efficacy ($\xi$) between the two groups (student's t-test). As shown in

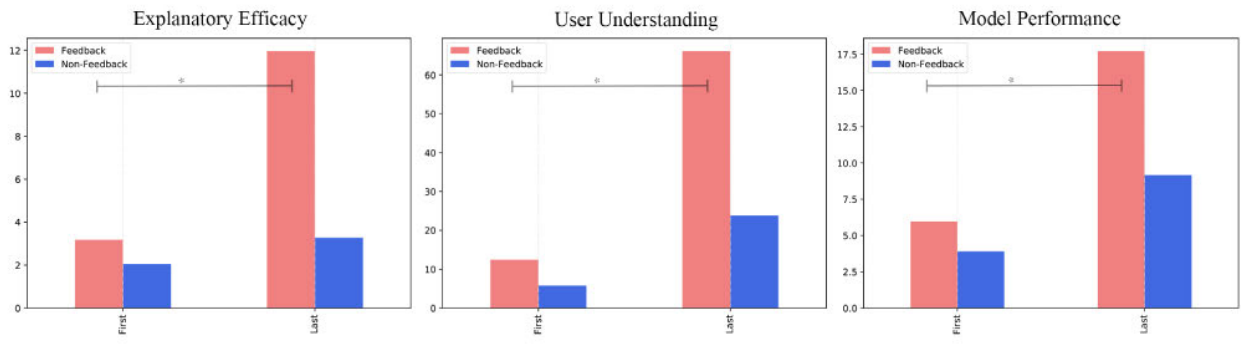

**FIGURE 7.** **The *feedback* group (Red) showed greater explanatory efficacy($\xi$) than interactions from the *non-feedback* group (Blue) during both the first five iterations and the last five iterations. * indicates the significant difference at the level $p < 0.01$.**

FIGURE 7, we found that explanatory efficacy in the *feedback* group had significantly higher scores than those in the *non-feedback* group for the last iteration ($p < 0.01, t = 4.43$). Changes in the scores (the last five – the first five iterations) of explanatory efficacy for the *feedback* group were greater than those for the *non-feedback* group ($p < 0.01, t = 3.85$). These results provide insights into aspects of the practicality of end-users comprehending and debugging the reasoning of an interactive system. The participants in the *feedback* group were significantly more likely to correctly and confidently agree with the machine-generated reasoning of the recommendation ($p < 0.01, t = 4.43, 3.85$). Explanatory efficacy increasingly improved during the iterative tasks in which participants interacted with the recommendation system. The feedback-based interaction helped the participants develop their understanding through memory retention and recall processes about its reasoning. In our post-task surveys across experiments, participants were generally positive about our interface:

"Compared to the first few lists, the recommendations that the system provided after a few trials were good and reflected my feedback. The more I used the system, the more I understood the explanations it provided and the more accurate the recommendations became."

However, repeated usage without any direct feedback on explanatory features (the *non-feedback* group) did not significantly improve their understanding ($p > 0.05$). Surprisingly, this result is not consistent with recent works in interactive machine learning, which have found that, for some systems, repeated use taught people the most salient aspects of how the system worked [34]. Such contrasting results support our hypothesis in this work. Interactive learning can ease users' understanding of a system only if the cyclic relationship between the decision-making process of a system and the user's feedback is well-established.

All together, these results imply that the participants in the *feedback* group had greater satisfaction and understanding of how they received the recommendations than those in the *non-feedback* group. The above findings provide evidence that allowing for end-users to give feedback by adjusting the agent's reasoning of an interactive system, such as correcting the system's flawed reasoning, can significantly improve the explanatory efficacy. In short, the explanatory process of how the system predicts the results can be delivered to users more effectively when users can provide feedback on the recommendations to the system.

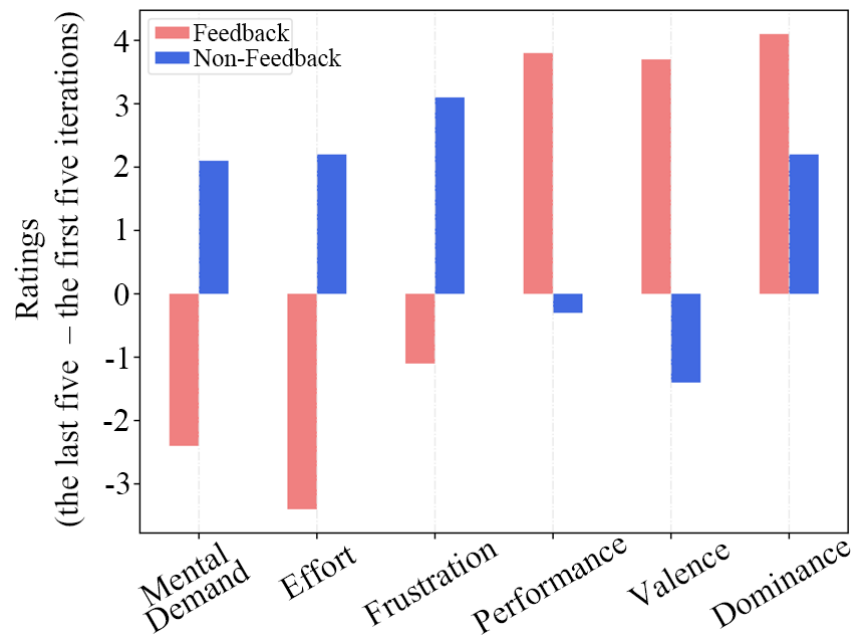

**FIGURE 8. Comparative results of the affective-cognitive questionnaires (the last five iterations – the first five iterations) between the two groups (Red – the *feedback* group, Blue – the *non-feedback* group). End-users of the *feedback* group perceived the explanatory process to be able to elicit more positive feelings and dominance in terms of controlling the recommendation system while requiring low cognitive workload.**

## VI. ANALYSIS OF EXPLANATORY EFFICACY AND SUBJECT RATINGS ON HUMAN AFFECT AND WORKLOAD

Participants perceived using the *non-feedback* explanatory interface to be more mentally demanding and more effort-inducing than using the *feedback* explanatory interface when performing the iterative tasks. FIGURE 8 shows the changes in the scores of the affective-cognitive elements of the questionnaire throughout the experiment for the two groups. They reported high scores in frustration, mental demand, and effort, failing to increase valence and overall performance even though the effort to understand the system increased. On the other hand, the participants in the *feedback* group improved in overall performance, valence, and dominance while mental demand and effort decreased. Dominance in both groups increased, but the *non-feedback* group showed higher levels than the other.

Furthermore, we analyzed the Pearson correlations between all components of the questionnaires. We note that we report and discuss effects that are significant with $p < 0.01$. As shown in Table 1, explanatory efficacy correlated highly positive with performance, valence and dominance, but moderately negative with mental demand and effort in the *feedback* group. In addition, we observed an improvement in performance with increased valence and dominance, but decreased mental demand for the participants in the *feedback* group as they performed the iterative tasks. In comparison, the *non-feedback* group showed the opposite trends. The users improved their overall performance when their work required greater mental demand and effort. However, the increase in effort failed to improve valence. Dominance in both groups have positive relationships with valence and effort, with the *feedback* group exhibiting a stronger relationship than the *non-feedback* group. These results indicate that the effects of explanatory efficacy on the perception of experience such as cognitive demands and emotional response are different for each user.

End-users perceived the explanatory process to elicit more positive feelings and dominance in terms of controlling the recommendation system while requiring low cognitive workload. It is possible that users had low cognitive workload in the *feedback* group because they were overwhelmed and had given up. The users may not understand how changing the weights would affect the recommendations, go through iterations without reasoning. These results are also corroborated by previous studies in which performance and controllability in mental models were discovered to be loosely connected [35]. For example, a few participants were overwhelmed by the abundant amount of information retrieved from the movie datasets:

"There are way too many features. Although the explanations tend to focus on a select few after a few trials, the vast amount of information I had to process in the beginning creates a learning curve, making it hard to understand the system well when first introduced to it."

However, mental demand in the *feedback* group showed negative correlations with both performance and dominance. Furthermore, performance and valence increased while effort decreased, and had no significant relationship with frustration. Thus, this evidence suggests that explanatory progress on how the system predicts the results does not overwhelm participants but instead gains confidence from users more effectively when users can provide feedback on the recommendations to the system.

**TABLE 1. A comparison between the two groups of the means of the group-wise intercorrelations ($p < 0.01$) between the scales of the cognitive (mental demand (M), effort (E), performance (P), frustration (F)) and affective (valence (V), dominance (D)) questionnaires. Explanatory progress on how the system predicts the results does not overwhelm participants but instead gains confidence from users more effectively when users can provide feedback on the recommendations to the system.**

| | | | *Feedback* | | | | |
|---|---|---|---|---|---|---|---|
| $\xi$ | M | E | P | V | D | F | |
| | | | -0.68 | | -0.28 | | F |
| 0.56 | -0.22 | 0.71 | | 0.62 | | -0.28 | D |
| 0.24 | | 0.59 | 0.51 | | 0.62 | | V |
| 0.73 | -0.61 | -0.39 | | 0.51 | | -0.68 | P |
| -0.32 | 0.34 | | -0.39 | 0.59 | 0.71 | | E |
| -0.18 | | 0.34 | -0.61 | | -0.22 | | M |
| | | | *Non-Feedback* | | | | |
| $\xi$ | M | E | P | V | D | F | |
| | | 0.54 | -0.32 | -0.25 | -0.22 | | F |
| 0.07 | -0.13 | 0.34 | 0.03 | 0.43 | | -0.22 | D |
| -0.12 | | | 0.33 | | 0.43 | -0.25 | V |
| 0.12 | 0.56 | 0.51 | | 0.33 | 0.03 | -0.32 | P |
| -0.38 | 0.69 | | 0.51 | | 0.34 | 0.54 | E |
| -0.23 | | 0.69 | 0.56 | | -0.13 | | M |

## VII. CORRELATES OF EXPLANATORY EFFICACY AND EEG

For the correlation statistic, we computed the $p$-values for the left and right-tailed correlation tests with the Spearman correlated coefficients. Assuming independence, this was done for each participant separately. The resulting $p$-values were then combined to one $p$-value via Fisher's method [36].

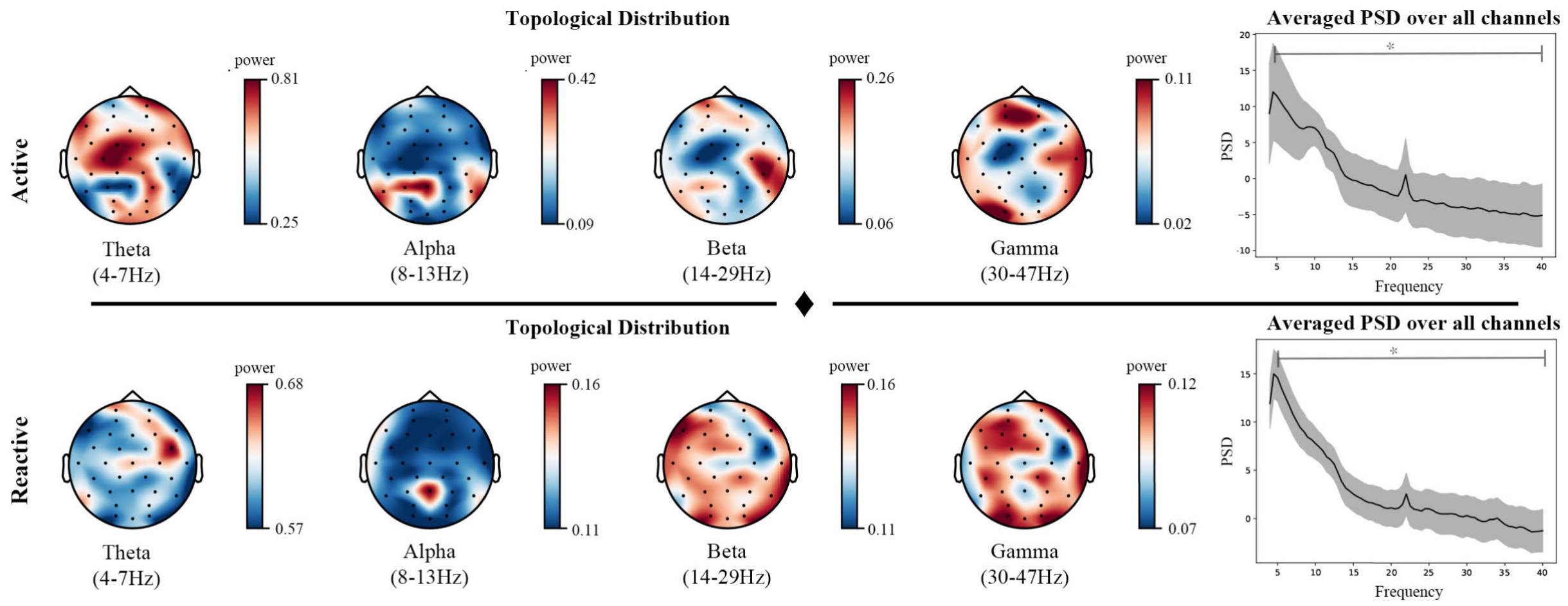

**FIGURE 9.** **Topographic distribution and averaged EEG power in the broad frequency bands of theta (4-7 Hz), alpha (8-13 Hz), beta (14-29 Hz), and gamma (30-47 Hz). Differences in neural activities are present when people are able to correct wrong predictions by the machine system and recognize their responses are being reflected in the learning mechanisms. The frequency in the frontal and temporal cortices was significantly activated when explanatory efficacy improved (*feedback* group). $*$ indicates the significant difference at the level $p < 0.01$, $t = -4.14$.**

**TABLE 2.** **Electrodes for which EEG signal was significantly correlated with explanatory efficacy ($p < 0.01$). Mean of the subject-wise correlations (R), the most negative (R−), and the most positive correlation (R+).**

| $\theta$ | | | | $\alpha$ | | | |
|---|---|---|---|---|---|---|---|
| Elec | R | R- | R+ | Elec | R | R- | R+ |
| C3 | 0.10 | -0.41 | 0.48 | P7 | 0.14 | -0.24 | 0.35 |
| Fc1 | 0.07 | -0.33 | 0.39 | P3 | 0.09 | -0.18 | 0.24 |
| Fp2 | -0.05 | -0.41 | 0.36 | Pz | 0.02 | -0.34 | 0.35 |
| F7 | 0.04 | -0.24 | 0.33 | | | | |
| $\beta$ | | | | $\gamma$ | | | |
| Elec | R | R- | R+ | Elec | R | R- | R+ |
| Cp6 | -0.17 | -0.61 | 0.43 | O1 | 0.12 | -0.14 | 0.25 |
| C4 | -0.11 | -0.47 | 0.32 | Fp1 | 0.09 | -0.04 | 0.12 |
| P8 | -0.05 | -0.21 | 0.23 | Fp2 | -0.04 | -0.12 | 0.19 |
| | | | | T8 | -0.02 | -0.09 | 0.05 |

The EEG signals of participants in the *feedback* group were significantly different from those in the *non-feedback* group. We analyzed the statistical difference in the mean changes of the four frequency bands of the EEG signals (FIGURE 9). The test revealed a significant difference in all frequency bands ($-0.34 \pm 3.93$ vs. $2.92 \pm 1.95$) of the conditions ($t = -4.14, p < 0.01$). We found that the signals had significantly different activations around the frontal and temporal lobes in most frequencies. The visualized result in FIGURE 9 implies neural activities are different when people are able to correct wrong predictions by the machine system and recognize their responses are being reflected in the learning mechanisms. We found that spatiotemporal characteristics of the brain can correlate to this perception associated with explanatory efficacy. The frequency in the frontal and temporal cortices was significantly activated when explanatory efficacy improved. A comprehensive list of the effects can be found in Table 2.

We found that EEG power in the theta, alpha and beta bands over frontal and temporal lobes had significant monotonic relationships in explanatory efficacy.

In light of the correlations between the affective-cognitive assessments on using our proposed interface (Table 2 and FIGURE 8), the greater the difference between the activations of the left hemisphere and the right hemisphere, the higher the explanatory efficacy, resulting in participants feeling more dominance. Furthermore, the left hemisphere showed greater activation than the right hemisphere in the *feedback* group. Many studies have discovered that the two halves of the frontal cortex are specialized with the left being involved in establishing positive feelings and the right half in establishing negative ones [25]. This implies that participants' feelings on dominance and explanatory efficacy could be characterized by EEG signals from the left half of the brain. This observation is consistent with the neuropsychological finding that the left side of the brain is responsible for being logical, analytical, and detail-oriented (FIGURE 2). The discovered monotonic relationship ($p < 0.01$) in the alpha and beta frequency bands supports neuroscientific studies on the effect of the alpha and beta bands on the frontal and temporal lobes in conscious thought, perception, emotions, and personality [22], [31]. Thus, the discovered spatiotemporal characteristics in the brain affect perceptions of explanatory efficacy and give us reason to investigate the potential of EEG signals as a measure for explanatory efficacy (Q3).

We found that analyzing asymmetry between the left and right hemispheres may be a possible way to evaluate the explanatory efficacy, with this method having support from well-known affective theories: the valence and withdrawal-approach hypotheses described in Section 2. We tested whether there is a significant difference in the mean changes of the lateralized powers as in [22] between the *feedback* and *non-feedback* groups. The results on the mean changes of the lateralization (See FIGURE 10) revealed a significant difference at all frequency bands ($p < 0.01$) on different conditions ($t = 27.68$, 11,08, 36.59, and 51.64). Hence, hemispheric asymmetry of EEG signals can be a measure of explanatory efficacy. If the alpha and beta frequencies of an end-user's left half of the brain increased more than those of the right hemisphere while interacting with an iterative machine system, we can determine that the user built a good understanding of the machine's behaviors and the system predicted well.

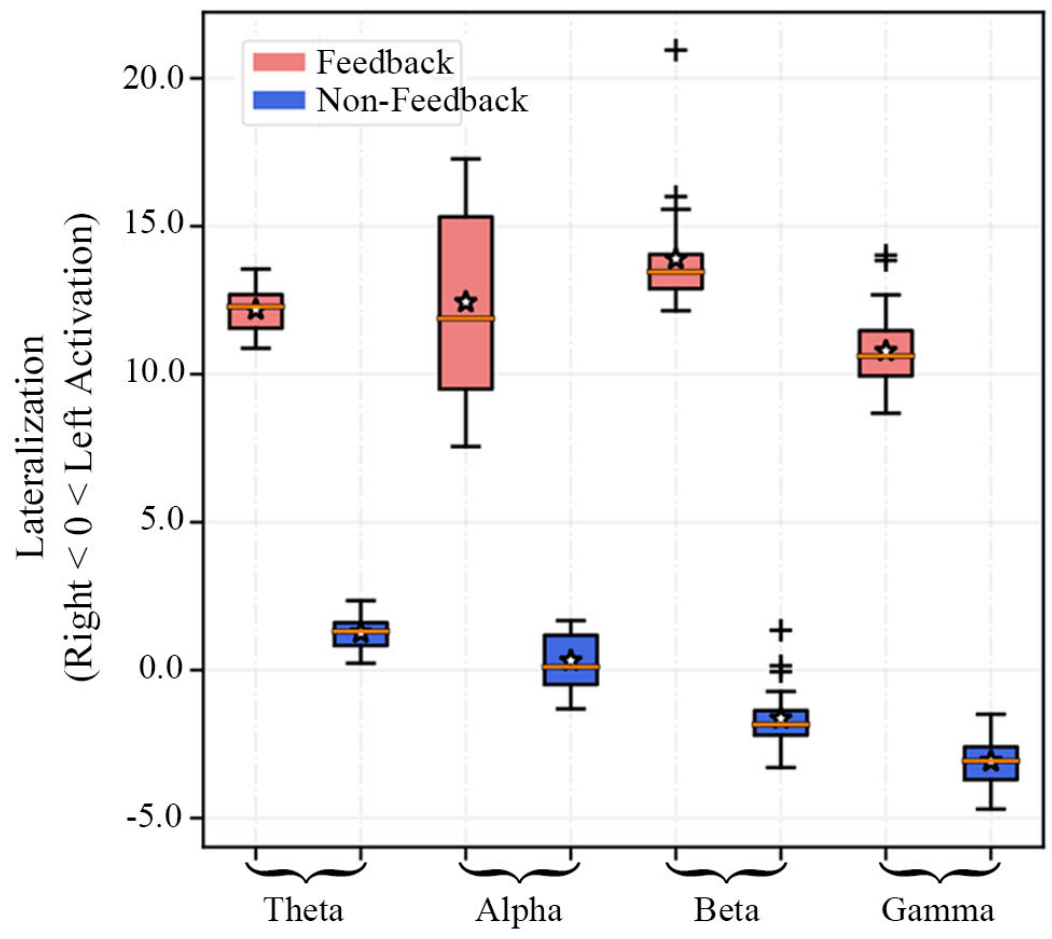

**FIGURE 10.** **The mean changes of the lateralized powers (theta, alpha, beta, and gamma) between the *feedback* and *non-feedback* groups. The left hemisphere exhibited greater activation compared to the right hemisphere when explanatory efficacy improved (*feedback* group).**

To further test EEG signals as an alternative measure for explanatory efficacy, we pose a single-trial EEG classification of the explanatory efficacy. We used EEG signals as inputs and the participants' ratings of explanatory efficacy as the ground truth. The three classes were determined by dividing the 9-point rating scale of the participants' explanatory efficacy into three classes (low, mid, and high), with each class containing three points. As the results from Table 2 and FIGURE 10 imply that analyzing asymmetry between the left and right hemispheres may be a possible way to evaluate explanatory efficacy, we extracted three different features from the EEG signals: hemispheric asymmetry (HASYM) [22], differential entropy (DE) [37], and asymmetric differential entropy (ADE) [38], which is the difference between the DEs of pairs of hemispheric asymmetry electrodes (FIGURE 5). For classifiers, the linear-kernel support vector machine (SVM) and $k$-nearest neighbors ($k$NN) algorithm were used after reducing the feature dimension by two components using the principal component analysis (PCA) algorithm. In each trial, we evaluate the performance of classification in terms of F1 score for each participant in a leave-one-out cross-validation scheme. At each step of the cross-validation, one trial was used as the test set, and the rest was used as a training set.

The averaged F1-score over participants resulted in Table 3. This result is consistent with the previous results shown in Table 2, and FIGURE 9 and 10. The explanatory efficacy related to EEG in beta frequency more closely than other frequency bands. Hemispheric features such as HASYM and ADE had more discriminative power to classify the explanatory efficacy. Altogether, the result indicates that we can measure and evaluate the efficacy of the machine's explanations with EEG signals. Further research can shed additional light on this potential by developing robust computational models to learn the characteristics of EEG signals on the hemispheric lateralization [22] such as those associated with explanatory efficacy.

**TABLE 3.** **Averaged classification accuracies (F1-Scores) using different features.**

| Feature | Classifier | Frequency Band $\theta$ | $\alpha$ | $\beta$ | $\gamma$ |
|---|---|---|---|---|---|
| HASYM | SVM | 0.566 | 0.533 | **0.611** | 0.477 |
| | $k$NN | 0.553 | **0.572** | 0.565 | 0.558 |
| DE | SVM | 0.487 | **0.495** | 0.493 | 0.455 |
| | $k$NN | 0.474 | 0.455 | **0.514** | 0.471 |
| ADE | SVM | 0.498 | 0.591 | **0.635** | 0.596 |
| | $k$NN | 0.512 | 0.587 | **0.624** | 0.600 |

## VIII. POTENTIALITY AND LIMITATION

Through answering the three questions, we have found that the EEG signals correlated to explanatory efficacy of users who were able to improve their understanding by providing *explnatory, interactive feedback* differ from those of users who were only able to provide *one-way explanation*. We have also discovered that asymmetric patterns in the left and right hemispheres depended on explanatory efficacy, supporting previously constructed neuroscientific theories. These findings indicate that we can measure and evaluate the efficacy of the machine's explanations with neural signals.

With the results from our exploration of the use of EEG in measuring the affective-cognitive processes of our participants, we suggest that using EEG is practical for the evaluation of the explanatory efficacy of interfaces that support analytical reasoning tasks. EEG is uniquely well suited for analyzing affective-cognitive processing in human behaviors, as its excellent temporal resolution allows for tracking neural responses in real-time. Such evaluation based on neural response can overcome some existing issues of self-assessments with validity and corroboration (e.g., participants may not answer with exactly how they are feeling, but instead give responses similar to those they expect others would likely provide). Furthermore, quantitative evaluation of how end-users perceive the machine's explanation is available without self-assessment tools. This advantage should be particularly appealing for interface designers, as traditional performance metrics cannot be applied to predictive models of great complexity.

Considering the machine learning perspective, we should also note that our iterative and interactive system is not limited to a movie recommendation system. It can be fully generalizable for other purposes. It only requires multi-variable data as input and task-specified parameters for optimizing predictions. The generalizability of our explanatory interface can allow for future researchers to develop interactive machine learning systems that reenact the machine decision-making process by enabling the human-AI partnership. For instance, a doctor-AI collaborative system that detects cancer from complex medical images can apply our interactive system to improve its accuracy. Given the images as inputs and typical shapes or sizes of tumors as labels, the AI can diagnose cancer and provide explanations about its decision-making process, i.e., suspected lesion of the specific image area as potential cancer. Doctors can then improve the AI's performance by correcting what they perceive to be the AI's flawed reasoning with a wearable EEG device. The implicit feedback from brain activities can replace explicit feedback, reducing the physical time it takes to correct the AI over numerous mispredictions. In such a symbiotic relationship, they can achieve a shared goal: cancer detection.

Therefore, we believe that our EEG-based evaluation of interactive machine learning systems can open new perspectives for integrating human-artificial intelligence to solve problems. The circular ecosystem in the partnership can make explicit representations, opening the "black boxes" in the machine's decision-making process. The "open boxes" in such learning approaches can further reinforce the role of human intelligence when using machine intelligence to discover relationships and solve problems.

## IX. CONCLUSION

We have demonstrated that the implementation and usage of our feedback-based explanation interface improved the users' affective-cognitive processes. Users could improve their understanding as they received explanations for the system's decisions and provide feedback on those decisions to the machine; simultaneously, the system succeeds in utilizing that feedback to achieve personalization with high precision. More importantly, we verified that the explanatory efficacy of the interface can be evaluated with EEG signals. The neuroscientific characteristics of EEG signals were correlated with affective-cognitive processes in perceptually driven iterative tasks and indicative of whether users realize a greater understanding of our system. We, therefore, suggest that EEG will lend insight into the evaluation of explanatory interfaces during complex analytical interactions.

Our future work will be developing a computational model to reduce the gap between human perception and the machine's ability. The complexity of neural mechanisms has often led to difficulties in measuring and accurately understanding affective-cognitive processes. EEG signals are affected by human thoughts and emotions are often subject to noise from various artifacts, low signal-to-noise ratio (SNR) of sensors, and inter- and intra- subject variability in physiological activation [24]. We believe building reliable automated systems for overcoming such challenging issues can accelerate the usability evaluation process in real-world applications.

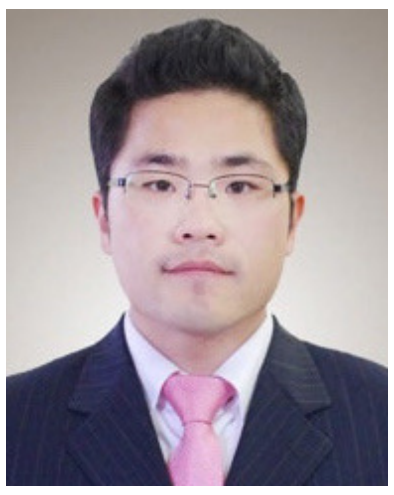

**BYUNG HYUNG KIM** received the B.S. degree in computer science from Inha University, South Korea, in 2008, the M.S. degree in computer science from Boston University, Boston, MA, USA, in 2010, and the Ph.D. degree in computer science from KAIST, South Korea, in 2018. Since 2018, he has been a Research Assistant Professor with the School of Computing, KAIST. His research interests include affective computing, brain–computer interface, computer vision, assistive and rehabilitative technology, and cerebral asymmetry and the effects of emotion on brain structure.

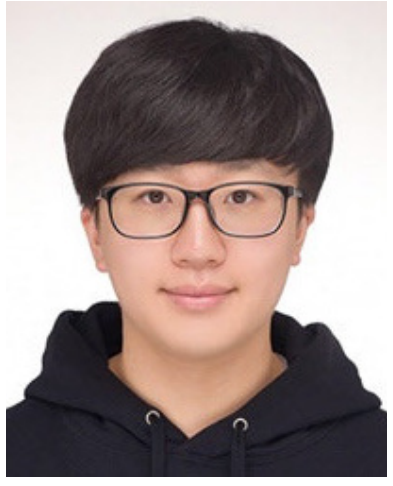

**SEUNGHUN KOH** received the B.S. degree in mathematical sciences from KAIST, Daejeon, South Korea, in 2019, where he is currently pursuing the M.S. degree in computer science. His research interests include explainable AI and human–computer interaction.

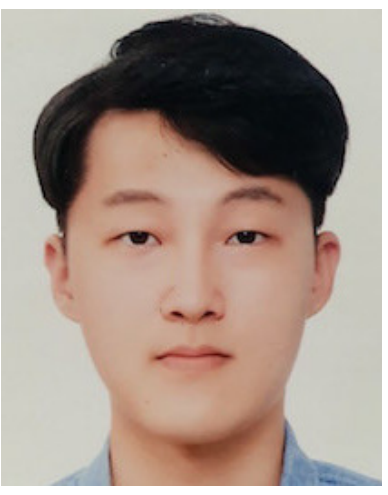

**SEJOON HUH** received the B.S. degree in computer science from KAIST, Daejeon, South Korea, in 2019, where he is currently pursuing the master's degree in computer science. His research interests include brain–computer interface applications and applied machine learning.

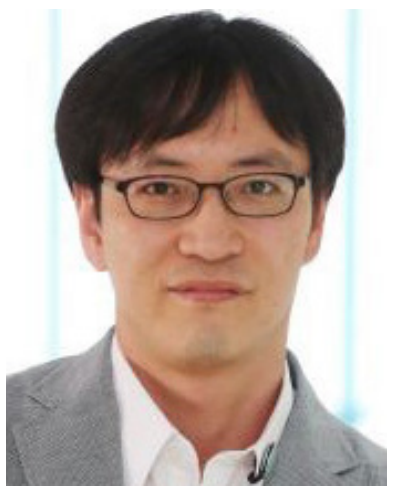

**SUNGHO JO** received the B.S. degree from the School of Mechanical and Aerospace Engineering, Seoul National University, Seoul, South Korea, in 1999, and the S.M. degree in mechanical engineering and the Ph.D. degree in electrical engineering and computer science from the Massachusetts Institute of Technology (MIT), Cambridge, MA, USA, in 2001 and 2006, respectively. While pursuing the Ph.D., he was associated with the Computer Science and Artificial Intelligence Laboratory (CSAIL), the Laboratory for Information Decision and Systems (LIDS), and Harvard-MIT HST Neuro Engineering Collaborative. Before joining the faculty at KAIST, he worked as a Postdoctoral Researcher at the MIT Media Lab. Since December 2007, he has been with the Department of Computer Science, KAIST, where he is currently a Professor. His research interests include intelligent robots, neural interfacing computing, and wearable computing.

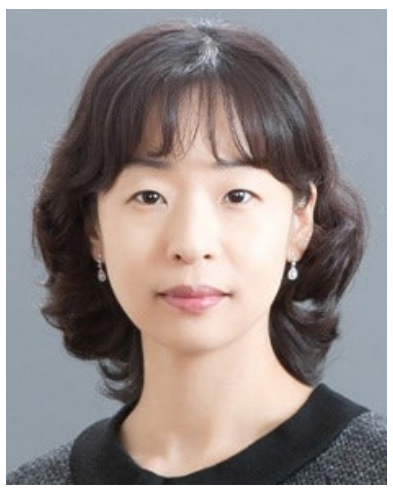

**SUNGHEE CHOI** received the B.S. degree in computer engineering from Seoul National University, in 1995, and the M.S. and Ph.D. degrees in computer science from The University of Texas at Austin, in 1997 and 2003, respectively. She is currently an Associate Professor of Computer Science with KAIST. Her research interests include geometric computing, computational geometry, geometric modeling, and computer graphics.